# Impact of Cooperation in Flow-Induced Diffusive Mobile Molecular Communication

Neeraj Varshney[†], Adarsh Patel[‡], Werner Haselmayr[§], Aditya K. Jagannatham[†], Pramod K. Varshney[‡], Weisi Guo[♯]

[†] Department of Electrical Engineering, Indian Institute of Technology Kanpur, Kanpur UP 208016, India
(e-mail:{neerajv; adityaj}@iitk.ac.in).

[‡] Department of Electrical Engineering & Computer Science, Syracuse University, Syracuse, NY 13244, USA
(e-mail:{apatel31; varshney}@syr.edu).

[§] Johannes Kepler University Linz, Austria (email: werner.haselmayr@jku.at).

[♯] School of Engineering, University of Warwick, United Kingdom (email: weisi.guo@warwick.ac.uk).

*Abstract*—Motivated by the numerous healthcare applications of molecular communication (MC) inside blood vessels, this work considers relay/cooperative nanomachine (CN)-assisted mobile MC between a source nanomachine (SN) and a destination nanomachine (DN) where each nanomachine is mobile in a flow-induced diffusive channel. Using the first hitting time model, the impact of an intermediate CN on the performance of the CN-assisted diffusive mobile MC system with fully absorbing receivers is analyzed in the presence of inter-symbol interference, multi-source interference, and counting errors. For this purpose, the likelihood ratio test based optimal symbol detection scheme is obtained at the DN considering the non-ideal nature of CN, i.e., CN can be in error with a finite probability. Further, to characterize the system performance, closed-form expressions for the end-to-end probabilities of detection and false alarm at the DN are derived between the SN-DN pair incorporating the detection performance of the intermediate CN. In addition, the channel capacity expression is also derived for the aforementioned scenario. Simulation results are presented to corroborate the theoretical results derived and also, to yield insights into system performance.

## I. INTRODUCTION

*Artificial molecular communication (MC)*, that focuses on the design, fabrication and testing of human-made MC systems, has led to the development of novel applications [1] such as efficient chrono drug-delivery [2] and human body monitoring, e.g., the detection and monitoring cholesterol or disease precursors in the blood vessels [3], using communicating nano-robots or nanomachines. To realize an efficient chrono drug-delivery system, a nanomachine in one part of the body senses an event and communicates to a drug-delivery nanomachine in another part of the body using molecular signaling via the circulation system with advection flow [2]. Fig. 1 shows a conceptual system diagram, whereby communicating nanomachines, separated by a significant distance in a blood circulatory system, undergo a combined diffusion-advection transport process. The diffusion-advection channels modelled here are relevant for low Peclet number regimes (1 or less, where mass diffusion dominates viscosity) and for low Womersley number regimes (1 or less, where mass diffusion dominates inertial forces from circulation). This typically occurs in capillaries far from the heart. However, in this scenario, molecular concentration decays inversely as the cube of the distance between the mobile transmitter and receiver nanomachines [4], which severely limits the performance of such systems. Relay or cooperative nanomachine (CN)-assisted communication, where a mobile CN cooperates with a mobile source nanomachine (SN) in forwarding information to a mobile destination nanomachine (DN) as shown in Fig. 1, has been shown to successfully overcome this impediment by significantly enhancing the communication range and detection performance. Several works in the existing literature have analyzed the performance of CN-assisted MC systems [5]–[10]. However, to the best of our knowledge, none of the works consider the CN-assisted MC system with fully absorbing receivers and nanomachine mobility in a flow-induced diffusive channel in the presence of inter-symbol interference (ISI), multi-source interference (MSI), and counting errors.

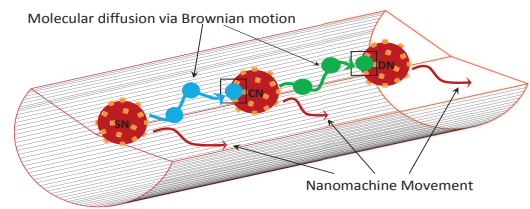

Fig. 1. CN-assisted mobile MC in diffusion-advection blood flow channels.

This work, therefore, analyzes the impact of an intermediate CN on the end-to-end detection performance of a CN-assisted mobile molecular communication system where mobile CN decodes the symbol with a finite error probability and forwards it to a mobile DN in the subsequent time-slot. The likelihood ratio test (LRT)-based optimal decision rule and the optimal decision threshold are determined at the mobile DN, incorporating the detection performance of intermediate mobile CN, in the presence of ISI, MSI, and counting errors for a flow-induced diffusive channel. Further, closed-form expressions are derived for the probabilities of detection and false alarm to analytically characterize the detection performance of the CN-assisted communication between the mobile SN and DN nodes. In addition, the channel capacity is also presented for the aforementioned scenario considering nanomachine mobility in a flow-induced diffusive channel.

## II. CN-ASSISTED DIFFUSIVE MOBILE MC SYSTEM MODEL

Consider a CN-assisted MC system where SN, CN, and DN are mobile with diffusion coefficients $D_{\text{sn}}, D_{\text{rn}}$, and $D_{\text{dn}}$, respectively, while communicating in a semi-infinite one-dimensional (1D) flow-induced channel with constant temperature and viscosity as illustrated in Fig. 1. The considered system model incorporates mobility to the system considered in [10]. The molecular propagation from the SN to CN and CN to DN occurs via Brownian motion with diffusion coefficient $D_{\text{p}}$. The communication channel is divided into time-slots of duration $\tau$ where the $j$th and $(j+1)$th slots are defined as the time intervals $[(j-1)\tau, j\tau]$ and $[j\tau, (j+1)\tau]$, respectively. In this system, the end-to-end communication between the SN and DN occurs in two time-slots. At the beginning of the $j$th time-slot, the SN emits either $\mathcal{Q}_0[j]$ number of type-0 molecules in the propagation medium for information symbol $x[j] = 1$ generated with a prior probability $\beta$ or remains silent for information symbol $x[j] = 0$. The intermediate CN decodes the symbol received from the SN, with probability of detection $(P_D^r[j])$ and probability of false alarm $(P_F^r[j])$, followed by the retransmission of $\mathcal{Q}_1[j+1]$ number of type-1 molecules or remains silent in the subsequent $(j+1)$th time-slot to indicate decoded symbols $\widehat{x}[j] = 1$ and $\widehat{x}[j] = 0$ respectively. Finally, the DN decodes the information symbol transmitted by the intermediate CN using the number of molecules received at the end of the $(j+1)$th time-slot.

Due to the stochastic nature of the diffusive channel, the times of arrival at the DN, of the molecules emitted by the CN, are random in nature and can span multiple time slots. Let $q_{j-i}$ denote the probability that a molecule transmitted by CN in slot $i \in \{1, 2, \cdots, j\}$ arrives at DN during time-slot $j$, which is given in [11, Eq. (1)], as $q_{j-i} = \int_{(j-i)\tau}^{(j-i+1)\tau} f(t;i)dt$, where $f(t;i)$ is the probability density function (PDF) of the first hitting time, i.e., the time required for a molecule to reach the DN. The PDF $f(t;i)$ for a flow-induced diffusive channel considering mobile CN and DN, is given by [12, Eq. (16)]

$$f(t;i) = \frac{\sqrt{i\tau D_{\text{tot}}D}}{\pi\sqrt{t}v(t;i)}\exp\left(-\frac{d_0^2}{4i\tau D_{\text{tot}}}\right) + \frac{d_0}{\sqrt{4\pi D(u(t;i))^3}}$$
$$\times \exp\left(-\frac{d_0^2}{4Du(t;i)}\right)\text{erf}\left(\frac{d_0}{2}\sqrt{\frac{tD}{i\tau D_{\text{tot}}v(t;i)}}\right), \quad (1)$$

where $u(t;i) \triangleq t + i\tau D_{\text{tot}}/D$ and $v(t;i) \triangleq i\tau D_{\text{tot}} + tD$. The distance $d_0$ is the Euclidean distance between the TN and RN, $\text{erf}(x)$ denotes the standard error function and the quantities $D_{\text{tot}}$ and $D$ are defined as, $D_{\text{tot}} = D_{\text{rn}} + D_{\text{dn}}$ and $D = D_{\text{dn}} + D_{\text{p}}$ respectively. The intermediate CN transmits $\mathcal{Q}_1[j+1]$ molecules for the decoded symbol $\widehat{x}[j] = 1$. The total number of type-1 molecules received $R_{rd}[j+1]$ at the DN during time-slot $[j\tau, (j+1)\tau]$ can be expressed as

$$R_{rd}[j+1] = S_{rd}[j+1] + \mathcal{I}_{rd}[j+1] + N_{rd}[j+1] + C_{rd}[j+1], \quad (2)$$

where $S_{rd}[j+1]$ is the number of type-1 molecules received from the CN during the current slot $[j\tau, (j+1)\tau]$ and follows Binomial distribution [13]. The quantity $N_{rd}[j+1]$ denotes MSI and follows a Gaussian PDF $N_{rd}[j+1] \sim \mathcal{N}(\mu_o, \sigma_o^2)$ [14] and the counting error $C_{rd}[j+1]$ follows a Gaussian PDF [15], given as $C_{rd}[j+1] \sim \mathcal{N}(0, \mathbb{E}\{R_{rd}[j+1]\})$. The ISI $\mathcal{I}_{rd}[j+1]$ at the DN at slot $j+1$ arises due to the transmission of previous $j-1$ slots, given as $\mathcal{I}_{rd}[j+1] = \sum_{i=2}^{j} I_{rd}[i]$, where $I_{rd}[i]$ follows a Binomial distribution. The noise $N_{rd}[j+1]$, the number of molecules $S_{rd}[j+1]$, $\mathcal{I}_{rd}[j+1]$ received from the intended TN are independent [10], [15], [16]. Further, for sufficiently large number of molecules released by the CN, the Binomial PDFs $S_{rd}[j+1]$ and $I_{rd}[i]$ can be approximated by the Gaussian PDFs as $S_{rd}[j+1] \sim \mathcal{N}(\mathcal{Q}_1[j+1]\widehat{x}[j]q_0, \mathcal{Q}_1[j+1]\widehat{x}[j]q_0(1-q_0))$ and $I_{rd}[i] \sim \mathcal{N}(\mathcal{Q}_1[j-i+2]\widehat{x}[j-i+1]q_{i-1}, \mathcal{Q}_1[j-i+2]\widehat{x}[j-i+1]q_{i-1}(1-q_{i-1}))$, $2 \leq i \leq j$.

## III. DETECTION PERFORMANCE ANALYSIS

The symbol detection problem in (2) at the DN for transmission of decoded symbols $\widehat{x}[j] = 0$ and $\widehat{x}[j] = 1$ by the intermediate CN in the $(j+1)$th time-slot can be formulated as the binary hypothesis testing problem

$$\begin{aligned}\mathcal{H}_0 &: R_{rd}[j+1] = \mathcal{I}_{rd}[j+1] + N_{rd}[j+1] + C_{rd}[j+1] \\ \mathcal{H}_1 &: R_{rd}[j+1] = S_{rd}[j+1] + \mathcal{I}_{rd}[j+1] + N_{rd}[j+1] \\ &\quad + C_{rd}[j+1],\end{aligned} \quad (3)$$

where $\mathcal{H}_0$ and $\mathcal{H}_1$ denote the null and alternative hypotheses corresponding to the transmission of decoded symbols $\widehat{x}[j] = 0$ and $\widehat{x}[j] = 1$ respectively by the intermediate CN during the $(j+1)$th time-slot. The quantity $R_{rd}[j+1]$ corresponding to the two hypotheses in (3) follows a distribution, given as

$$\begin{aligned}\mathcal{H}_0 &: R_{rd}[j+1] \sim \mathcal{N}(\mu_0[j+1], \sigma_0^2[j+1]) \\ \mathcal{H}_1 &: R_{rd}[j+1] \sim \mathcal{N}(\mu_1[j+1], \sigma_1^2[j+1]),\end{aligned} \quad (4)$$

where the mean $\mu_0[j+1]$ and the variance $\sigma_0^2[j+1]$ under the null hypothesis $\mathcal{H}_0$ are calculated as

$$\mu_0[j+1] = \beta\sum_{i=2}^{j}\mathcal{Q}_1[j-i+2]q_{i-1} + \mu_o, \quad (5)$$

$$\sigma_0^2[j+1] = \sum_{i=2}^{j}\{\beta\mathcal{Q}_1[j-i+2]q_{i-1}(1-q_{i-1}) + \beta(1-\beta) \\ \times (\mathcal{Q}_1[j-i+2]q_{i-1})^2\} + \sigma_o^2 + \mu_0[j+1]. \quad (6)$$

Similarly, mean $\mu_1[j+1]$ and variance $\sigma_1^2[j+1]$ corresponding to the alternative hypothesis $\mathcal{H}_1$ are derived as

$$\mu_1[j+1] = \mathcal{Q}_1[j+1]q_0 + \mu_0[j+1], \quad (7)$$
$$\sigma_1^2[j+1] = \mathcal{Q}_1[j+1]q_0(2-q_0) + \sigma_0^2[j+1]. \quad (8)$$

The optimal decision rule at the DN is obtained next.

*Theorem 1: The LRT based optimal detector at the DN corresponding to transmission by the intermediate CN at the $(j+1)$th time-slot is obtained as*

$$T(R_{rd}[j+1]) = R_{rd}[j+1] \underset{\mathcal{H}_0}{\overset{\mathcal{H}_1}{\gtrless}} \gamma[j+1], \quad (9)$$

*where the optimal decision threshold $\gamma[j+1]$ is computed as $\gamma[j+1] = \sqrt{\gamma'[j+1]} - \alpha[j+1]$ with $\alpha[j+1]$ and $\gamma'[j+1]$ defined as*

$$\alpha[j+1] = \frac{\mu_1[j+1]\sigma_0^2[j+1] - \mu_0[j+1]\sigma_1^2[j+1]}{\sigma_1^2[j+1] - \sigma_0^2[j+1]}, \quad (10)$$

$$\gamma'[j+1] = \zeta[j+1]\frac{2\sigma_1^2[j+1]\sigma_0^2[j+1]}{\sigma_1^2[j+1] - \sigma_0^2[j+1]} + (\alpha[j+1])^2$$
$$+ \frac{\mu_1^2[j+1]\sigma_0^2[j+1] - \mu_0^2[j+1]\sigma_1^2[j+1]}{\sigma_1^2[j+1] - \sigma_0^2[j+1]}, \quad (11)$$

where
$$\zeta[j+1] = \ln\left[\frac{\sigma_1[j+1]\{(1-\beta)(1-P_F^r[j]) - \beta(1-P_D^r[j])\}}{\sigma_0[j+1]\{\beta P_D^r[j] - (1-\beta)P_F^r[j]\}}\right], \quad (12)$$

that incorporates the detection performance $(P_D^r[j], P_F^r[j])$ of intermediate CN.

**Proof** See Appendix A. □

The result below determines the resulting detection performance at the DN in the CN-assisted diffusive MC network.

*Theorem 2:* The average probabilities of detection $P_D$ and false alarm $P_F$ at the DN corresponding to SN transmissions in slots $1$ to $k$ for the CN-assisted flow-induced diffusive MC system with nanomachine mobility are given as

$$P_D = \frac{1}{k}\sum_{j=1}^{k} P_D^d[j+1], \quad (13)$$

$$P_F = \frac{1}{k}\sum_{j=1}^{k} P_F^d[j+1], \quad (14)$$

where the probabilities of detection $P_D^d[j+1]$ and false alarm $P_F^d[j+1]$ at the DN in the $(j+1)$th slot are obtained as

$$P_D^d[j+1] = Q\left(\frac{\gamma[j+1] - \mu_1[j+1]}{\sigma_1[j+1]}\right) P_D^r[j]$$
$$+ Q\left(\frac{\gamma[j+1] - \mu_0[j+1]}{\sigma_0[j+1]}\right)(1 - P_D^r[j]), \quad (15)$$
$$P_F^d[j+1] = Q\left(\frac{\gamma[j+1] - \mu_1[j+1]}{\sigma_1[j+1]}\right) P_F^r[j]$$
$$+ Q\left(\frac{\gamma[j+1] - \mu_0[j+1]}{\sigma_0[j+1]}\right)(1 - P_F^r[j]), \quad (16)$$

where $Q(\cdot)$ denotes the Q-function.

**Proof** See Appendix B. □

## IV. CAPACITY ANALYSIS

Let the discrete random variables $X[j]$ and $Y[j+1]$ represent the transmitted and received symbol in the $j$th and $(j+1)$th slots respectively. The mutual information $I(X[j], Y[j+1])$ between $X[j]$ and $Y[j+1]$ for the CN-assisted link can be expressed as

$$I(X[j], Y[j+1])$$
$$= \Pr(y[j+1]=0|x[j]=0)\Pr(x[j]=0)$$
$$\times \log_2 \frac{\Pr(y[j+1]=0|x[j]=0)}{\sum_{x[j]\in\{0,1\}} \Pr(y[j+1]=0|x[j])\Pr(x[j])}$$
$$+ \Pr(y[j+1]=0|x[j]=1)\Pr(x[j]=1)$$
$$\times \log_2 \frac{\Pr(y[j+1]=0|x[j]=1)}{\sum_{x[j]\in\{0,1\}} \Pr(y[j+1]=0|x[j])\Pr(x[j])}$$
$$+ \Pr(y[j+1]=1|x[j]=0)\Pr(x[j]=0)$$
$$\times \log_2 \frac{\Pr(y[j+1]=1|x[j]=0)}{\sum_{x[j]\in\{0,1\}} \Pr(y[j+1]=1|x[j])\Pr(x[j])}$$
$$+ \Pr(y[j+1]=1|x[j]=1)\Pr(x[j]=1)$$
$$\times \log_2 \frac{\Pr(y[j+1]=1|x[j]=1)}{\sum_{x[j]\in\{0,1\}} \Pr(y[j+1]=1|x[j])\Pr(x[j])}, \quad (17)$$

where $\Pr(x[j]=1)=\beta$, $\Pr(x[j]=0)=1-\beta$, and $\Pr(y[j+1] \in \{0,1\}|x[j]\in\{0,1\})$ can be determined in terms of $P_D^d[j+1]$ and $P_F^d[j+1]$ as

$$\Pr(y[j+1]=0|x[j]=0) = 1 - P_F^d[j+1],$$
$$\Pr(y[j+1]=1|x[j]=0) = P_F^d[j+1],$$
$$\Pr(y[j+1]=0|x[j]=1) = 1 - P_D^d[j+1],$$
$$\Pr(y[j+1]=1|x[j]=1) = P_D^d[j+1].$$

The capacity $C[k]$ of the CN-assisted diffusive channel, as $k$ approaches $\infty$ [11], is now obtained by maximizing the mutual information $I(X[j], Y[j+1])$ for $1 \leq j \leq k$ as

$$C[k] = \max_{\beta} \frac{1}{(k+1)} \sum_{j=1}^{k} I(X[j], Y[j+1]) \text{ bits/slot}. \quad (18)$$

The factor $\frac{1}{k+1}$ arises in the above expression due to the fact that $k+1$ time-slots are required to communicate $k$ bits from the SN to DN in the CN-assisted system.

## V. SIMULATION RESULTS

Fig. 2 demonstrates the detection performance at the DN in the CN-assisted diffusive mobile MC system under various scenarios. For simulation purposes, the probabilities of detection $P_D^r[j]$ and false alarm $P_F^r[j]$ at the CN are fixed as mentioned in Figs. 2a and 2b. The other parameters are set as, diffusion coefficient $D_p = 5 \times 10^{-10}$ m$^2$/s, slot duration $\tau = 10$ ms, prior probability $\beta = 0.5$, drift velocity $v=10^{-3}$ m/s, $k = 10$ slots, and Euclidean distance $d_0 = 1\mu$m. The MSI at the DN is modeled as a Gaussian distributed RV with mean $\mu_o = 10$ and variance $\sigma_o^2 = 10$.

First, it can be observed from Fig. 2a that the analytical values derived in (13) and (14) coincide with those obtained from simulations, thus validating the analytical results. Further, the detection performance at the DN significantly improves with the improvement in performance of intermediate CN. However, the detection performance at the DN saturates on further improvement in performance of intermediate CN. This is due to the fact that the end-to-end performance of the CN-assisted system is dominated by the weak CN-DN link. Fig. 2b shows that the system performs identically in a fluidic medium with and without drift. This is owing to the fact that the arrival probabilities $q_{j-i}$ are equivalent under both the scenarios. One can also observe that an increase in the number of molecules $\mathcal{Q}_1[j+1]$ emitted by the CN for symbol $\widehat{x}[j]=1$ results in a higher probability of detection at the DN for a fixed value of probability of false alarm. For example, as $\mathcal{Q}_1[j+1]$ increases from 30 to 60 the probability of detection $P_D$ at the DN significantly improves from 0.5 to 0.8 at probability of false alarm $P_F = 0.01$ for a scenario when the DN is mobile with diffusion coefficient $D_{\text{dn}} = 5 \times 10^{-13}$ m$^2$/s. However, the detection performance deteriorates as the diffusion coefficient $D_{\text{dn}}$ increases due to higher mobility of the DN.

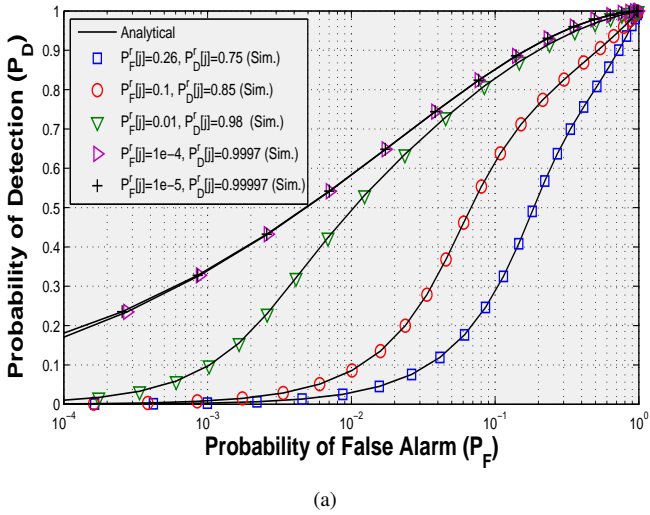
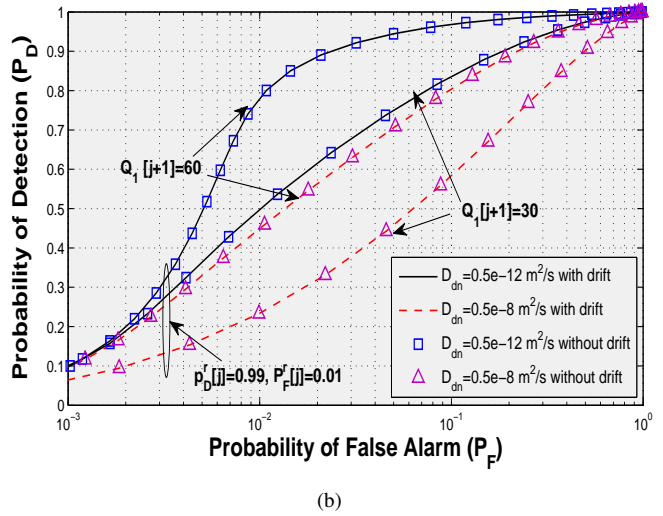

(a)  (b)

Fig. 2. Detection performance of a CN-assisted diffusive mobile MC system with (a) $\mathcal{Q}_1[j+1]=30 \; \forall \; \widehat{x}[j]=1$, $D_{\text{rn}}=10^{-10}$ m$^2$/s, and $D_{\text{dn}}=0.5\times10^{-12}$ m$^2$/s (b) $\mathcal{Q}_1[j+1]\in\{30, 60\} \; \forall \; \widehat{x}[j] = 1$, and $D_{\text{rn}}=10^{-10}$ m$^2$/s.

Fig. 3 shows the capacity performance of the CN-assisted flow-induced diffusive mobile MC system under various scenarios. As depicted in Fig. 3, the capacity of MC decreases significantly as the variance ($\sigma_o^2$) of the MSI increases. Further, one can also observe that the capacity of the CN-assisted system depends substantially on the detection performance of the intermediate CN. As the detection performance $(P_D^r[j], P_F^r[j])$ at the CN increases from $(0.85, 0.1)$ to $(0.99, 0.01)$, a significant capacity gain can be achieved at low as well as high MSI. However, similar to detection performance at the DN, the channel capacity decreases as the diffusion coefficient $D_{\text{dn}}$ increases due to higher mobility of the DN.

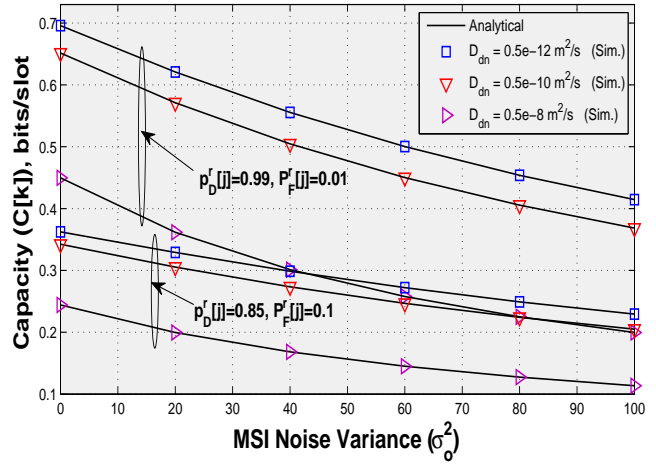

Fig. 3. Capacity versus MSI variance ($\sigma_o^2$) with $\mathcal{Q}_1[j+1] = 60 \; \forall \; \widehat{x}[j]=1$, $D_{\text{rn}}=10^{-10}$ m$^2$/s, and $\mu_o = 10$.

## APPENDIX A
## PROOF OF THEOREM 1

The optimal LR based test $\mathcal{L}(R_{rd}[j+1])$ at the mobile DN is acquired as

$$\mathcal{L}(R_{rd}[j+1]) = \frac{p(R_{rd}[j+1]|\mathcal{H}_1)}{p(R_{rd}[j+1]|\mathcal{H}_0)} \underset{\mathcal{H}_0}{\overset{\mathcal{H}_1}{\gtrless}} \frac{1-\beta}{\beta}, \quad (19)$$

and further evaluated in (20). Using the PDFs $p(R_{rd}[j+1]|\widehat{x}[j] = 1)$ and $p(R_{rd}[j+1]|\widehat{x}[j] = 0)$, given as

$$p(R_{rd}[j+1]|\widehat{x}[j] = 1)$$
$$= \frac{1}{\sqrt{2\pi\sigma_1^2[j+1]}} \exp\left\{-\frac{(R_{rd}[j+1] - \mu_1[j+1])^2}{2\sigma_1^2[j+1]}\right\}, \quad (21)$$

$$p(R_{rd}[j+1]|\widehat{x}[j] = 0)$$
$$= \frac{1}{\sqrt{2\pi\sigma_0^2[j+1]}} \exp\left\{-\frac{(R_{rd}[j+1] - \mu_0[j+1])^2}{2\sigma_0^2[j+1]}\right\}, \quad (22)$$

in (20) and taking logarithm after cross multiplication, (20) is expressed as

$$\frac{(R_{rd}[j+1]-\mu_0[j+1])^2}{2\sigma_0^2[j+1]}-\frac{(R_{rd}[j+1]-\mu_1[j+1])^2}{2\sigma_1^2[j+1]} \underset{\mathcal{H}_0}{\overset{\mathcal{H}_1}{\gtrless}} \zeta[j+1],$$

which can be further simplified to yield

$$\frac{1}{2\sigma_0^2[j+1]\sigma_1^2[j+1]}f(R_{rd}[j+1]) \underset{\mathcal{H}_0}{\overset{\mathcal{H}_1}{\gtrless}} \zeta[j+1], \quad (23)$$

$$\mathcal{L}(R_{rd}[j+1]) = \frac{p(R_{rd}[j+1]|\widehat{x}[j]=1)\Pr(\widehat{x}[j]=1|\mathcal{H}_1) + p(R_{rd}[j+1]|\widehat{x}[j]=0)\Pr(\widehat{x}[j]=0|\mathcal{H}_1)}{p(R_{rd}[j+1]|\widehat{x}[j]=1)\Pr(\widehat{x}[j]=1|\mathcal{H}_0) + p(R_{rd}[j+1]|\widehat{x}[j]=0)\Pr(\widehat{x}[j]=0|\mathcal{H}_0)}$$
$$= \frac{p(R_{rd}[j+1]|\widehat{x}[j]=1)P_D^r[j] + p(R_{rd}[j+1]|\widehat{x}[j]=0)(1-P_D^r[j])}{p(R_{rd}[j+1]|\widehat{x}[j]=1)P_F^r[j] + p(R_{rd}[j+1]|\widehat{x}[j]=0)(1-P_F^r[j])}. \quad (20)$$

where $\zeta[j+1]$ and $f(R_{rd}[j+1])$ are defined in (12) and (24), respectively. Further, the expression for $f(R_{rd}[j+1])$ in (23) is simplified as follow

$$f(R_{rd}[j+1]) \triangleq (R_{rd}[j+1] - \mu_0[j+1])^2 \sigma_1^2[j+1] \\ - (R_{rd}[j+1] - \mu_1[j+1])^2 \sigma_0^2[j+1] \quad (24)$$
$$= R_{rd}^2[j+1](\sigma_1^2[j+1] - \sigma_0^2[j+1]) \\ + 2R_{rd}[j+1](\mu_1[j+1]\sigma_0^2[j+1] - \mu_0[j+1]\sigma_1^2[j+1]) \\ + (\mu_0^2[j+1]\sigma_1^2[j+1] - \mu_1^2[j+1]\sigma_0^2[j+1])$$
$$= (\sigma_1^2[j+1] - \sigma_0^2[j+1])(R_{rd}[j+1] + \alpha[j+1])^2 \\ - (\mu_1[j+1]\sigma_0^2[j+1] - \mu_0[j+1]\sigma_1^2[j+1])\alpha[j+1] \\ + (\mu_1[j+1]\sigma_0^2[j+1] - \mu_0[j+1]\sigma_1^2[j+1]). \quad (25)$$

Using $f(R_{rd}[j+1])$ from (25) in (23) and solving for the number of received molecules $R_{rd}[j+1]$ the test reduces to

$$(R_{rd}[j+1] + \alpha[j+1])^2 \underset{\mathcal{H}_0}{\overset{\mathcal{H}_1}{\gtrless}} \gamma'[j+1], \quad (26)$$

where $\alpha[j+1]$ and $\gamma'[j+1]$ are defined in (10) and (11). Taking the square root in (26), where $\gamma'[j+1] \geq 0$ and $\alpha[j+1] \geq 0$ that can be readily seen from (7) and (8), yields the optimal test in (9) at the DN.

## APPENDIX B
## PROOF OF THEOREM 2

The detection probability $P_D^d[j+1]$ at the DN can be derived using the decision rule in (9) as

$$P_D^d[j+1] = \Pr(T(R_{rd}[j+1]) > \gamma[j+1]|\mathcal{H}_1)$$
$$= \Pr(R_{rd}[j+1] > \gamma[j+1]|\widehat{x}[j]=1)\Pr(\widehat{x}[j]=1|\mathcal{H}_1) \\ + \Pr(R_{rd}[j+1] > \gamma[j+1]|\widehat{x}[j]=0)\Pr(\widehat{x}[j]=0|\mathcal{H}_1)$$
$$= \Pr(R_{rd}[j+1] > \gamma[j+1]|\widehat{x}[j]=1)P_D^r[j] \\ + \Pr(R_{rd}[j+1] > \gamma[j+1]|\widehat{x}[j]=0)(1-P_D^r[j]). \quad (27)$$

Similarly, the false alarm probability $P_F^d[j+1]$ can be obtained as,

$$P_F^d[j+1] = \Pr(T(R_{rd}[j+1]) > \gamma'_{sd}[j+1]|\mathcal{H}_0)$$
$$= \Pr(R_{rd}[j+1] > \gamma'_{rd}[j+1]|\widehat{x}[j]=1)\Pr(\widehat{x}[j]=1|\mathcal{H}_0) \\ + \Pr(R_{rd}[j+1] > \gamma'_{rd}[j+1]|\widehat{x}[j]=0)\Pr(\widehat{x}[j]=0|\mathcal{H}_0)$$
$$= \Pr(R_{rd}[j+1] > \gamma'_{rd}[j+1]|\widehat{x}[j]=1)P_F^r[j] \\ + \Pr(R_{rd}[j+1] > \gamma'_{rd}[j+1]|\widehat{x}[j]=0)(1-P_F^r[j]). \quad (28)$$

Further, substituting

$$\Pr(R_{rd}[j+1] > \gamma[j+1]|\widehat{x}[j]=1) \\ = Q\left(\frac{\gamma[j+1] - \mu_1[j+1]}{\sigma_1[j+1]}\right) \quad (29)$$

and

$$\Pr(R_{rd}[j+1] > \gamma[j+1]|\widehat{x}[j]=0) \\ = Q\left(\frac{\gamma[j+1] - \mu_0[j+1]}{\sigma_0[j+1]}\right) \quad (30)$$

in (27) and (28), one can obtain the final expressions for $P_D^d[j+1]$ and $P_F^d[j+1]$ given in (15) and (16) respectively.